\documentclass[prl, twocolumn, showpacs, superscriptaddress, floatfix]{revtex4}
\usepackage{epsfig}
\usepackage{amsmath}
\usepackage{graphicx}
\usepackage{bbm}
\usepackage[usenames]{color}

\begin{document}
\author{I. E. Linington}
\affiliation{Department of Physics and Astronomy, University of Sussex,
  Falmer, Brighton, BN1 9QH, United Kingdom}
\affiliation{Department of Physics, Sofia University, James Bourchier 5 blvd, 1164 Sofia,
Bulgaria}
\author{N. V. Vitanov}
\affiliation{Department of Physics, Sofia University, James Bourchier 5 blvd, 1164 Sofia,
Bulgaria}
\affiliation{Institute of Solid State Physics, Bulgarian Academy of Sciences,
Tsarigradsko chauss\'{e}e 72, 1784 Sofia, Bulgaria}
\title{Robust creation of arbitrary-sized Dicke states of trapped ions by global addressing}
\begin{abstract}
We propose a novel technique for the creation of maximally entangled symmetric Dicke states in an ion trap using adiabatic passage, which requires only a pair of chirped pulses from a single laser and is applicable to any number of ions and excitations. By utilising a particular factorisation of the Hilbert space for multi-level ladders we show that the problem can be reduced to `bow-tie' configuration energy-level crossings. This technique is naturally robust against fluctuations in the laser intensity and the chirp rate. Even when realistic heating rates are considered, we estimate that the overall fidelity should remain high (e.g. $98\%$ for a $10$-ion Dicke state), which represents a significant improvement over traditional approaches.
\end{abstract}
%
%
\maketitle
{\bf \em Introduction}. 
Genuine multipartite entangled states involving many excitations are centrally important for quantum information science. They are essential for quantum communication across more than one channel, provide a universal resource for quantum computation and can be used to improve the sensitivity of atomic clocks. Furthermore, developing a clear understanding of entanglement is fundamental to our description of microscopic systems. However, both the production and characterisation of {\it{large}} entangled states remain formidable challenges.
A particularly important class of multi-particle entangled states are the Dicke-symmetric states of $N$ particles and $m$ excitations:
\begin{align}
\left\vert{}W_{m}^{N}\right\rangle & \equiv \frac{1}{\sqrt{C_{m}^{N}}}\sum_{k}P_{k}\vert{}\underbrace{1,1,\ldots,1}_{m\;\textrm{excitations}},0,\ldots,0\rangle,
\end{align}
where $\left\{P_{k}\right\}$ denotes the set of all distinct combinations
of ions \cite{mandel1995} and \mbox{$C_{m}^{N}\equiv{}N!/\left[m!(N-m)!\right]$}. 
These states are robust against decoherence, particle loss and measurement. Also, the entanglement that they contain cannot be destroyed by local operations \cite{kiesel2007}.

Currently, the maturest experimental approach to the creation of large entangled states is to couple trapped ions by allowing them to interact with a common vibrational mode. This can be achieved either with resonant laser pulses of precise area \cite{cirac1995, haffner2005} or off-resonant geometric gates \cite{sorensen1999, leibfried2005}, and Dicke states containing a single quantum of energy (the so-called $W$-states) have been fabricated in chains of up to 8 ions using the first technique \cite{haffner2005}. However, addressing the ions one or two at a time requires a costly increase in the number of preparation steps as the complexity of the final state grows, and for this reason, Dicke-symmetric states with more than one quantum of energy represent a significant theoretical and experimental challenge. 
The creation of Dicke-states has been proposed using simultaneous addressing for a coupled spin-chain, by using many precisely-timed pulses to navigate through a network of energy-level crossings \cite{unayan2002}.
In this Letter, we propose an {\it{extremely simple}} method for producing $\left\vert{}W_{m}^{N}\right\rangle$ states robustly, for arbitrary $m$ and $N$ using a two-step adiabatic-passage technique.

\begin{figure}
\includegraphics[width=0.9\columnwidth]{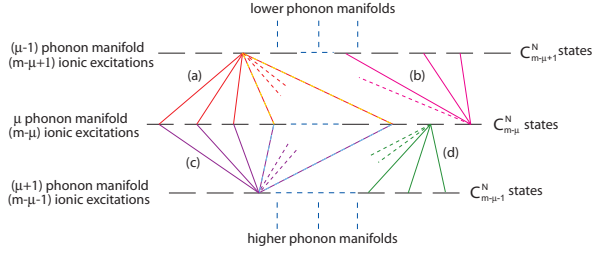}
\caption{Level-scheme for $N$ ions coupled to their common centre-of-mass mode and sharing $m$ quanta. 
(The detuning is not shown in this figure.) 
The number of quanta is conserved by (\ref{Hamiltonian}) and so dynamics within the subspace containing $m$ excitations is closed. This subspace consists of $(m+1)$ separate manifolds, 
each coupled only to its neighbours.}
\label{level_scheme_fig}
\end{figure}

{\bf \em Level-scheme}.
We consider $N$ ions cooled and trapped in a linear array. Each ion has two relevant internal states, $\vert0\rangle$ and $\vert1\rangle$, and interacts with a laser pulse that is tuned close to (but not exactly at) the frequency of the first red sideband of the lowest vibrational mode. 
We assume that the phonon spectrum can be resolved sufficiently well that only the levels of the centre of mass mode are excited by this interaction and that other vibrational modes can safely be neglected. In the Lamb-Dicke limit, the interaction Hamiltonian is ($\hbar=1$) \cite{cirac1995}:
\begin{align}
\widehat{H}_{I} & = \sum_{j=1}^{N}\frac{\eta\Omega_{j}'(t)}{2\sqrt{N}}\Bigg[\vert{}1\rangle_{j}\langle{}0\vert_{j}\hat{a}\exp\left(-i\int_{0}^{t}\Delta_{j}(\tau)\;d\tau\right)
\nonumber \\ &
\hspace{16mm}+\vert{}0\rangle_{j}\langle{}1\vert_{j}\hat{a}^{\dagger}\exp\left(+i\int_{0}^{t}\Delta_{j}(\tau)\;d\tau\right)\Bigg],
\label{Hamiltonian}
\end{align}
where $\Omega_{j}' $ is the Rabi-frequency for the $j^{th}$ ion, $\eta$ is the single-ion Lamb-Dicke parameter, $\Delta_{j}$ is the $j^{th}$ ion-laser detuning, and $N$ the number of ions. Equation (\ref{Hamiltonian}) holds in the limit that the trap frequency, \mbox{$\nu_{\textrm{tr}}^{2}\gg(2.6\Omega_{j}(t)\eta)^{2}/N$} \cite{cirac1995, james1998}.
Since the total number of excitations, 
\mbox{$\widehat{N} = \hat{a}^{\dagger}\hat{a} + \sum_{j=1}^{N}\vert{}1\rangle_{j}\langle{}1\vert_{j}$},
is conserved, the dynamics within the subspace corresponding to a given number of quanta is closed. It is sufficient to consider the case of $m$-excitations shared collectively between the $N$ ions and the bus mode \cite{cirac1995}. This subspace splits nicely into manifolds corresponding to $\mu=0,1,\ldots,m$ phonons in the bus-mode, each of which is ($C_{m-\mu}^{N}$)-fold degenerate. The relevant level scheme is sketched in figure \ref{level_scheme_fig}.
%
%
\begin{figure}[!htb]
\includegraphics[width=0.75\columnwidth]{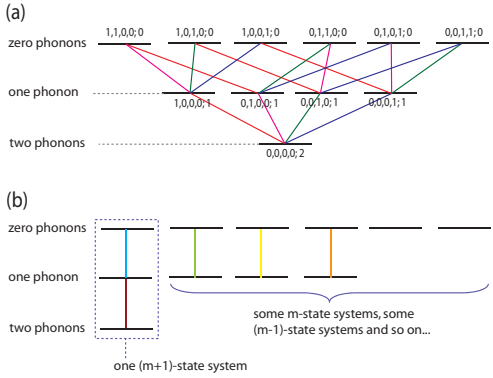}
\caption{Illustration of the Morris-Shore factorisation for $N=4$ and $m=2$.
(a) Level-scheme for 4 ions and 2 excitations. 
Different colour lines represent the coupling between the bus mode and the four different ions. 
(b) The same system in the Morris-Shore basis consists of a single \mbox{$(m+1)$}-level subsystem, some $m$-level subsystems, some \mbox{$(m-1)$}-level subsystems etc, all the way down to a collection of dark states. By ensuring that the required initial and final states are confined to the $(m+1)$-level subspace, the dynamics can be simplified considerably, since the coupling between different subsystems is zero.}
\label{MS_4_ions}
\end{figure}

{\bf \em Morris-Shore factorisation}.
Even though the total number of energy quanta is preserved by the Hamiltonian, (\ref{Hamiltonian}), the relevant Hilbert space contains $\sum_{\mu=0}^{m}C_{\mu}^{N}$ states, which grows very rapidly as either the number of ions or the number of excitations is increased. This means that an exact analytic treatment of this system rapidly becomes unfeasible using conventional approaches. From a practical perspective, the number of control steps required to navigate between different states of this large Hilbert space soon becomes prohibitively high, since the overall fidelity is degraded with each control step. 
Fortunately, a very convenient transformation exists which greatly simplifies the problem \cite{rangelov2006, morris1983}.  By making a suitable change of basis, it is possible to de-couple the relevant state space into a series of smaller subspaces. These subspaces are not connected by  the Hamiltonian and so the dynamics within them remains closed. This procedure is known as the Morris-Shore factorisation, and is sketched schematically in figure \ref{MS_4_ions}. Where it exists, this factorisation can provide a remarkable simplification of the dynamics of many-level systems. However, in the general case, its existence is not guaranteed, since the couplings between the neighbouring manifolds are required to satisfy a specific commutation relation. 

Below, we shall show that for the present system the Morris-Shore transformation \emph{does} exist, and furthermore it is possible to confine the dynamics to a single ladder involving $(m+1)$-levels. 
By inducing a suitable level crossing within this subspace it is possible to transfer population adiabatically between the lowest and highest states of this $(m+1)$-level chain. In this way, Dicke-symmetric states may be robustly created in a single step for arbitrary $N$ and $m$.
The laser pulses we shall choose involve the same Rabi frequency $\Omega'(t)$ and detuning $\Delta(t)$ for all ions, and for ease of notation, we define the effective Rabi-frequency \mbox{$\Omega(t)\equiv\eta\Omega_{j}'(t)/\sqrt{N}$}. 
Experimentally this is an appealing choice, since this interaction Hamiltonian may be realised by shining a single laser pulse onto all of the $N$ ions.
We proceed by changing the interaction picture to one in which the Hamiltonian can be written in block-matrix form:
\begin{eqnarray}
\nonumber
\widehat{H}_{I} & =
\begin{pmatrix}
\Delta_{m} & V_{m,m-1} & 0 & \cdots & 0& 0
\\
V_{m,m-1}^{\dagger} & \Delta_{m-1} & V_{m-1,m-2} & \cdots & 0 & 0
\\
0 & V_{m-1,m-2}^{\dagger} & \Delta_{m-2} & \cdots & 0 & 0
\\
\vdots & \vdots & \vdots & \ddots & \vdots & \vdots
\\
0 & 0 & 0 & \cdots & \Delta_{1} & V_{1,0}
\\
0 & 0 & 0 & \cdots & V_{1,0}^{\dagger} & \Delta_{0} 
\end{pmatrix}
\label{matrix_Hamiltonian}
\end{eqnarray}

The detuning-blocks $\Delta_{\mu}(t)$ are proportional to the identity operator of dimension $C_{m-\mu}^{N}$ (which we shall label $\mathbbm{1}_{\mu}$), with equal elements $\left(2\mu-m\right)\Delta(t)/2$. The block $V_{\mu, \mu-1}(t)$ represents the coupling between the manifolds with $\mu$ and $(\mu-1)$ phonons.
The multi-level Morris-Shore transformation exists if and only if the commutation relation
\mbox{$\left[V_{\mu,\mu-1}(t)V_{\mu,\mu-1}^{\dagger}(t), V_{\mu+1,\mu}^{\dagger}(t)V_{\mu+1,\mu}(t)\right] = 0$} holds for all the intermediate manifolds, \mbox{(i.e. $\mu=1,\ldots,m-1$)} \cite{rangelov2006}. 

Now, since $\widehat{H}_{I}(t)$ is symmetric with respect to interchange of any two ions, the size of the non-zero elements in any one coupling block are the same. The magnitude of the non-zero elements in $V_{\mu, \mu-1}(t)$ is $\Omega(t)\sqrt{\mu}/2$. Therefore, we define scaled coupling blocks, \mbox{$\tilde{V}_{\mu, \mu-1} = 2V_{\mu, \mu-1}(t)/\Omega(t)\sqrt{\mu}$}, which are all time-independent. The elements of these scaled coupling blocks can only be 0 or 1. The above commutation relation can now be re-written as follows:
\begin{align}
\left[\tilde{V}_{\mu,\mu-1}\tilde{V}_{\mu,\mu-1}^{\dagger}, \tilde{V}_{\mu+1,\mu}^{\dagger}\tilde{V}_{\mu+1,\mu}\right] & = 0.
\label{scaled_commutation_relation}
\end{align}

{\it{Properties of the coupling blocks.}} There are four relevant groups of coupling, sketched in figure \ref{level_scheme_fig}:
(a) Going from the $(\mu-1)$-phonon to the $\mu$-phonon manifold involves converting an ionic excitation into a phonon. Since each state in the $(\mu-1)$-phonon manifold contains $(m-\mu+1)$ excited ions, each upper state is connected to this many lower states. 
(b) Coupling from the $\mu$-phonon to the $(\mu-1)$-phonon manifold involves converting a phonon into an ionic excitation. Since each state in the $\mu$-phonon manifold contains $(N-m+\mu)$ unexcited ions, each lower state is connected to $(N-m+\mu)$ upper states. 
Similarly, there are $(N-m+\mu+1)$ couplings in group (c) and $(m-\mu)$ in group (d).
The products $\tilde{V}_{\mu,\mu-1}\tilde{V}_{\mu,\mu-1}^{\dagger}$
can now be computed. The diagonal elements are \mbox{$(N-m+\mu)$}, while
the values of the off-diagonal elements of $\tilde{V}_{\mu,\mu-1}\tilde{V}_{\mu,\mu-1}^{\dagger}$ are simply the number of common states in the $(\mu-1)$-phonon manifold coupled to both of the corresponding states in the $\mu$-phonon manifold. (This is represented schematically by the red/yellow dashed lines in figure \ref{level_scheme_fig}.) These values can only ever be one (and there are $(m-\mu)(N-m+\mu)$ of these per row) or zero. For example, the states $\vert\ldots10\ldots\rangle\vert\mu\rangle$ and $\vert\ldots01\ldots\rangle\vert\mu\rangle$ are both coupled to the state $\vert\ldots11\ldots\rangle\vert\mu-1\rangle$, but there are no other states to which they can both be coupled by $\tilde{V}_{\mu,\mu-1}$.
A similar analysis can be used to calculate the matrix $\tilde{V}_{\mu+1,\mu}^{\dagger}\tilde{V}_{\mu+1,\mu}$. Once again, the values of these off diagonal elements are one if these two states differ by the position of a single ionic excitation (blue/pink dashed lines in figure \ref{level_scheme_fig}) and zero otherwise. These off-diagonal elements are the same as in  $\tilde{V}_{\mu,\mu-1}\tilde{V}_{\mu,\mu-1}^{\dagger}$, while
the diagonal elements are $(m-\mu)$, giving:
\begin{align}
\tilde{V}_{\mu,\mu-1}\tilde{V}_{\mu,\mu-1}^{\dagger} & = \tilde{V}_{\mu+1,\mu}^{\dagger}\tilde{V}_{\mu+1,\mu} + (N-2m+2\mu)\mathbbm{1}_{\mu}.
\label{coupling_relation}
\end{align}
Clearly, the matrices $\tilde{V}_{\mu,\mu-1}\tilde{V}_{\mu,\mu-1}^{\dagger}$ and $ \tilde{V}_{\mu+1,\mu}^{\dagger}\tilde{V}_{\mu+1,\mu}$ commute, and so the generalised Morris-Shore transformation always exists for $m$ excitations shared collectively between $N$ ions and the phonon mode.

{\it{Morris-Shore states.}} Since there is only one $m$-phonon state, there is only one $(m+1)$-level ladder in the Morris-Shore factorisation, and as we are interested in adiabatically coupling the lowest and highest states of this chain, only these two states are actually relevant. However, in order to be sure which of the Morris-Shore eigenstates in the zero-phonon manifold belongs at the top of the ladder, it is necessary to calculate all $(m+1)$ levels.
These are uniquely defined by the condition that  for each $\mu=1,\ldots,m$, the eigenstates of $\tilde{V}_{\mu,\mu-1}\tilde{V}_{\mu,\mu-1}^{\dagger}$ and $\tilde{V}_{\mu,\mu-1}^{\dagger}\tilde{V}_{\mu,\mu-1}$ (in neighbouring manifolds) \emph{have the same eigenvalue} \cite{rangelov2006}. 
After noting that the sum of the elements in any row of either of these two matrices is equal to $(m-\mu+1)(N-m+\mu)$, we conclude that the completely symmetric states, $\left\vert{}W_{m-\mu}^{N}\right\rangle\vert\mu\rangle$ (with $\mu=0,\ldots,m$) are the eigenstates of the $(m+1)$-level chain in the Morris-Shore basis. This subspace is de-coupled from all other states during the dynamics and, conveniently, the state that we would like to create, i.e. $\left\vert{}W_{m}^{N}\right\rangle$, is at the top of this ladder. 

We note that the only states involved in the dynamics are eigenstates of the combined ionic pseudospin, $\hat{J}_{z} = \frac{1}{2}\sum_{j=1}^{N}\left(\vert{}1\rangle_{j}\langle{}1\vert_{j}-\vert0\rangle_{j}\langle0\vert_{j}\right)$. This is a necessary feature, which is made clear by writing the Hamiltonian in the form $\widehat{H}_{I}(t)\propto\left(\hat{J}_{+}\hat{a}e^{i\varphi}+\textrm{h.c.}\right)$, with $\hat{J}_{+}=\sum_{j=1}^{N}\vert{}1\rangle_{j}\langle0\vert_{j}$ \cite{mandel1995}.
Below, we show how it is possible to use adiabatic techniques to steer the system into this target state. 

\begin{figure}[htb]
\includegraphics[width=0.8\columnwidth]{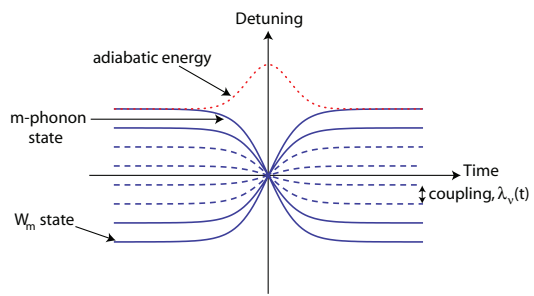}
\caption{The $m$-phonon state and the $\left\vert{}W_{m}^{N}\right\rangle$  state can be smoothly connected adiabatically using a level-crossing, providing that the couplings are chosen to satisfy the adiabaticity condition, (\ref{adiabaticity_condition}).}
\label{level_crossing_scheme}
\end{figure}

{\bf{\em Adiabatic creation of Dicke-symmetric states}}.
From now on, we shall only be concerned with the $(m+1)$-level ladder in the Morris-Shore factorisation.  The Rabi-frequency $\Omega(t)$ and detuning $\Delta(t)$ are selected in order to satisfy the objective of a smooth transfer of population between the two ends of this chain. Since the detunings, $\Delta_{\mu}(t)$, are all proportional to $\Delta(t)$, chirping the laser across resonance will lead to a `bow-tie' model, where all of the energies cross at the same instant of time.
The relevant level scheme is  shown in figure \ref{level_crossing_scheme}.
Analytic bow-tie solutions have been found for three \cite{carroll1986b} and $m$
states \cite{ostrovsky1997}. 
The couplings between each pair of successive Morris-Shore states in the Dicke ladder with $\mu$ and $\mu-1$ phonons are given by
\mbox{$\lambda_{\mu,\mu-1}(t)  = \frac{\Omega(t)}{2}\sqrt{\mu(m-\mu+1)(N-m+\mu)}$}.
We choose to use a `complex-sech' pulse, i.e.
\mbox{$\Omega(t)  = \Omega_{0}\textrm{sech}(t/T)$} and
\mbox{$\Delta(t)  = \Delta_{0}\textrm{tanh}(t/T)$}.
Compared to other types of level-crossing (e.g. Landau-Zener), these couplings result in a relatively fast adiabatic passage, with a very appropriate excitation profile; the excitation probability very close to unity in a certain range (controlled by $\Delta_{0}$) around resonance, and vanishing exponentially rapidly outside this range \cite{roos2004}. This is particularly relevant here, since it is important that our laser pulse does not excite unwanted vibrational modes. 
A transfer efficiency of $1-\epsilon$ between the zero-phonon and $m$-phonon states (with $\epsilon\ll1$) requires the coupling strength and detuning to satisfy
\begin{align}
\frac{(\pi \eta \Omega_0T)^{2}}{2N\ln(1/\epsilon)}\gtrsim \pi\Delta _{0}T\gtrsim m\ln(1/\epsilon).
\label{adiabaticity_condition}
\end{align}

The Dicke-symmetric state, $\left\vert{}W_{m}^{N}\right\rangle$, may be created as follows: (i) The system is prepared in its ground state and $m$ of the ions (arbitrarily chosen) are promoted from $\left\vert0\right\rangle$ to $\left\vert1\right\rangle$. This initialisation step may be performed robustly, using adiabatic techniques \cite{wunderlich2007}. (ii) A chirped laser pulse is focused onto the $m$ excited ions, creating an effective $m$-ion system containing $m$ quanta. The other $(N-m)$ ions remain in their ground states and de-couple from the dynamics. Providing the adiabaticity condition, (\ref{adiabaticity_condition}), is satisfied, the system will be transferred  smoothly into the state $\vert0\ldots0\rangle\vert{}m\rangle$.
Thus our scheme also provides a method for adiabatically creating number states of the phonon field, which will be described in detail elsewhere \cite{fock2008}. (Alternatively, this preparation may be achieved by performing stages (i) and (ii) on a single ion, $m$ times.) (iii) Finally, the laser beam is broadened in space to couple equally to all $N$ ions. By applying a second chirped pulse, the system can be transferred smoothly into the target state, $\vert{}W_{m}^{N}\rangle$.

\begin{figure}[!htb]
\includegraphics[width=0.83\columnwidth]{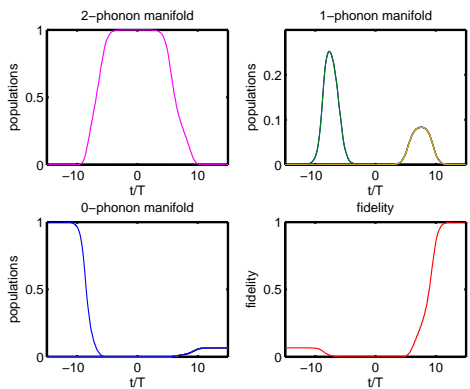}
\caption{The first three frames show the evolution of the populations of all 22 states for the creation of a $\left\vert{}W_{2}^{6}\right\rangle$ state. In the frames labelled 1-phonon manifold and 0-phonon manifold, several indistinguishable curves are overlaid, since certain ions are equivalent. At time $t\sim8T$, the state $\left\vert{}W_{1}^{6}\right\rangle$ is partially populated in the 1-phonon manifold. The bottom-right frame, labelled `fidelity', shows the time-dependent overlap with the required final state (final fidelity = $99.996\%$). Parameters are $\Omega_{0}T=10; \Delta_{0}T=6$.}
\label{6_ions_populations}
\end{figure}

Figure \ref{6_ions_populations} shows the result of a numerical simulation for the case of \mbox{$N=6, m=2$}.
 During the period \mbox{$-10T\leq{}t\leq0$}, population is transferred from state $\vert{}1,1,0,0,0,0\rangle\vert0\rangle$ to state $\vert{}0,0,0,0,0,0\rangle\vert2\rangle$ via the 1-phonon manifold by addressing just 2 ions. After $t=0$, the laser beam is spatially-broadened to couple equally to all 6 ions and the system is transferred back down into the 0-phonon manifold and into the symmetric state $\left\vert{}W_{2}^{6}\right\rangle$.
Since our approach uses adiabatic passage to navigate between states,
it is robust against fluctuations in the intensity and the timing of the laser pulses. Furthermore, even when intensity of the pulse was allowed to fluctuate by 10\% across the chain, the overall fidelity remained above 99.3\% for the parameters in figure \ref{6_ions_populations}. 

{\bf{\em{Heating effects.}}} 
$\Omega(t)$ is bounded from above by approximately
$\nu_{\textrm{tr}}/10$, in order to prevent extraneous excitations in
vibrational modes other than the centre of mass mode \cite{james1998}. Together with the adiabaticity requirement, (\ref{adiabaticity_condition}), this confines the
timescale, $T$ for the chirped pulses to be at least $25\mu{}s$ for a trap-frequency of 4MHz (as in \cite{schmidtkaler2003}). A
conservative estimate for the time required to implement the entire
technique is therefore $400\mu{}s$ (i.e. $16T$).
Taking a phonon heating-rate of $5s^{-1}$ per ion \cite{schmidtkaler2003} we therefore estimate 
that for a $10$-ion Dicke-state, the total amount of heating during the whole procedure could be on the level of $0.02$
phonons; the infidelity
introduced by bus-mode heating is thus of order $2\%$, which is significantly lower than could be expected from traditional approaches involving many sequential operations. Moreover, the effects of heating are largely insensitive to the complexity the target state.

%

{\bf{\em Conclusions}}.
We have shown that maximally-entangled Dicke states, $\vert{}W_{m}^{N}\rangle$, of arbitrary size may be generated in an ion-trap using a pair of extremely simple laser pulses. Starting with $m$ ions in state $\vert{}1\rangle$ and all other ions in state $\vert{}0\rangle$, the first chirped pulse simultaneously addresses these $m$ ions to create adiabatically a phonon number state with zero ionic excitations and $m$ phonons. A second chirped pulse addresses all $N$ ions to create the state $\vert{}W_{m}^{N}\rangle$.
This extremely simple scenario derives from three theoretical findings: (i) the multi-state Morris-Shore factorisation de-couples the vast Hilbert space of the $N$ ionic qubits into independent smaller chainwise linkages; (ii) the longest of these, an $(m+1)$-dimensional chain, happens to contain all the Dicke states from $\vert{}W_{1}^{N}\rangle$ to $\vert{}W_{m}^{N}\rangle$, along with the phonon-number state,  $\vert{}W_{0}^{N}\rangle$; (iii) using a frequency-chirped pulse creates a bow-tie linkage pattern, wherein the state  $\left\vert{}W_{0}^{N}\right\rangle$ connects adiabatically to the $\vert{}W_{m}^{N}\rangle$ state.
The proposed technique is adiabatic in nature and hence it is robust against intensity and frequency imperfections; moreover, because only two interaction steps are required, heating effects are reduced.
We thank Peter Ivanov and Peter Blythe for helpful discussions. This work has been supported by the EU ToK project CAMEL and the EU RTN project EMALI.


\begin{thebibliography}{15}
\expandafter\ifx\csname natexlab\endcsname\relax\def\natexlab#1{#1}\fi
\expandafter\ifx\csname bibnamefont\endcsname\relax
  \def\bibnamefont#1{#1}\fi
\expandafter\ifx\csname bibfnamefont\endcsname\relax
  \def\bibfnamefont#1{#1}\fi
\expandafter\ifx\csname citenamefont\endcsname\relax
  \def\citenamefont#1{#1}\fi
\expandafter\ifx\csname url\endcsname\relax
  \def\url#1{\texttt{#1}}\fi
\expandafter\ifx\csname urlprefix\endcsname\relax\def\urlprefix{URL }\fi
\providecommand{\bibinfo}[2]{#2}
\providecommand{\eprint}[2][]{\url{#2}}

\bibitem[{\citenamefont{Mandel and Wolf}(1995)}]{mandel1995}
\bibinfo{author}{\bibfnamefont{L.}~\bibnamefont{Mandel}} \bibnamefont{and}
  \bibinfo{author}{\bibfnamefont{E.}~\bibnamefont{Wolf}},
  \emph{\bibinfo{title}{Optical Coherence and Quantum Optics}}
  (\bibinfo{publisher}{Cambridge University Press}, \bibinfo{year}{1995}).

\bibitem[{\citenamefont{{Kiesel et~al}}(2007)}]{kiesel2007}
\bibinfo{author}{\bibfnamefont{N.}~\bibnamefont{{Kiesel et~al}}},
  \bibinfo{journal}{Phys. Rev. Lett.} \textbf{\bibinfo{volume}{98}},
  \bibinfo{pages}{063604} (\bibinfo{year}{2007}).

\bibitem[{\citenamefont{Cirac and Zoller}(1995)}]{cirac1995}
\bibinfo{author}{\bibfnamefont{J.~I.} \bibnamefont{Cirac}} \bibnamefont{and}
  \bibinfo{author}{\bibfnamefont{P.}~\bibnamefont{Zoller}},
  \bibinfo{journal}{Phys. Rev. Lett.} \textbf{\bibinfo{volume}{74}},
  \bibinfo{pages}{4091} (\bibinfo{year}{1995}).

\bibitem[{\citenamefont{{Haffner et~al}}(2005)}]{haffner2005}
\bibinfo{author}{\bibfnamefont{H.}~\bibnamefont{{Haffner et~al}}},
  \bibinfo{journal}{Nature} \textbf{\bibinfo{volume}{438}},
  \bibinfo{pages}{643} (\bibinfo{year}{2005}).

\bibitem[{\citenamefont{{S{\o}rensen} and {M{\o}lmer}}(1999)}]{sorensen1999}
\bibinfo{author}{\bibfnamefont{A.}~\bibnamefont{{S{\o}rensen}}}
  \bibnamefont{and}
  \bibinfo{author}{\bibfnamefont{K.}~\bibnamefont{{M{\o}lmer}}},
  \bibinfo{journal}{Phys. Rev. Lett.} \textbf{\bibinfo{volume}{82}},
  \bibinfo{pages}{1971} (\bibinfo{year}{1999}).

\bibitem[{\citenamefont{{Leibfried et~al}}(2005)}]{leibfried2005}
\bibinfo{author}{\bibfnamefont{D.}~\bibnamefont{{Leibfried et~al}}},
  \bibinfo{journal}{Nature} \textbf{\bibinfo{volume}{438}},
  \bibinfo{pages}{639} (\bibinfo{year}{2005}).

\bibitem[{\citenamefont{{Unanyan et~al}}(2002)}]{unayan2002}
\bibinfo{author}{\bibfnamefont{R.~G.} \bibnamefont{{Unanyan et~al}}},
  \bibinfo{journal}{Phys. Rev. A} \textbf{\bibinfo{volume}{66}},
  \bibinfo{pages}{042101} (\bibinfo{year}{2002}).

\bibitem[{\citenamefont{{James}}(1998)}]{james1998}
\bibinfo{author}{\bibfnamefont{D.~F.~V.} \bibnamefont{{James}}},
  \bibinfo{journal}{Appl. Phys. B} \textbf{\bibinfo{volume}{66}},
  \bibinfo{pages}{181} (\bibinfo{year}{1998}).

\bibitem[{\citenamefont{Rangelov et~al.}(2006)\citenamefont{Rangelov, Vitanov,
  and Shore}}]{rangelov2006}
\bibinfo{author}{\bibfnamefont{A.~A.} \bibnamefont{Rangelov}},
  \bibinfo{author}{\bibfnamefont{N.~V.} \bibnamefont{Vitanov}},
  \bibnamefont{and} \bibinfo{author}{\bibfnamefont{B.~W.} \bibnamefont{Shore}},
  \bibinfo{journal}{Phys. Rev. A} \textbf{\bibinfo{volume}{74}},
  \bibinfo{pages}{053402} (\bibinfo{year}{2006}).

\bibitem[{\citenamefont{Morris and Shore}(1983)}]{morris1983}
\bibinfo{author}{\bibfnamefont{J.~R.} \bibnamefont{Morris}} \bibnamefont{and}
  \bibinfo{author}{\bibfnamefont{B.~W.} \bibnamefont{Shore}},
  \bibinfo{journal}{Phys. Rev. A} \textbf{\bibinfo{volume}{27}},
  \bibinfo{pages}{906} (\bibinfo{year}{1983}).

\bibitem[{\citenamefont{Carroll and Hioe}(1986)}]{carroll1986b}
\bibinfo{author}{\bibfnamefont{C.~E.} \bibnamefont{Carroll}} \bibnamefont{and}
  \bibinfo{author}{\bibfnamefont{F.~T.} \bibnamefont{Hioe}},
  \bibinfo{journal}{J. Phys. A: Math. Gen.} \textbf{\bibinfo{volume}{19}},
  \bibinfo{pages}{2061} (\bibinfo{year}{1986}).

\bibitem[{\citenamefont{Ostrovsky and Nakamura}(1997)}]{ostrovsky1997}
\bibinfo{author}{\bibfnamefont{V.}~\bibnamefont{Ostrovsky}} \bibnamefont{and}
  \bibinfo{author}{\bibfnamefont{H.}~\bibnamefont{Nakamura}},
  \bibinfo{journal}{J. Phys. A} \textbf{\bibinfo{volume}{30}},
  \bibinfo{pages}{6939} (\bibinfo{year}{1997}).

\bibitem[{\citenamefont{Roos and Molmer}(2004)}]{roos2004}
\bibinfo{author}{\bibfnamefont{I.}~\bibnamefont{Roos}} \bibnamefont{and}
  \bibinfo{author}{\bibfnamefont{K.}~\bibnamefont{Molmer}},
  \bibinfo{journal}{Phys. Rev. A} \textbf{\bibinfo{volume}{69}},
  \bibinfo{pages}{022321} (\bibinfo{year}{2004}).

\bibitem[{\citenamefont{{Wunderlich et~al}}(2007)}]{wunderlich2007}
\bibinfo{author}{\bibfnamefont{C.}~\bibnamefont{{Wunderlich et~al}}},
  \bibinfo{journal}{J. Mod. Opt.} \textbf{\bibinfo{volume}{54}},
  \bibinfo{pages}{1541} (\bibinfo{year}{2007}).
  
\bibitem{fock2008}
\bibinfo{author}{\bibfnamefont{I.~E.} \bibnamefont{Linington et~al}},
\bibinfo{journal}{arXiv:0802.3538v1}
  (\bibinfo{year}{2008}).

\bibitem[{\citenamefont{{Schmidt-Kaler et~al}}(2003)}]{schmidtkaler2003}
\bibinfo{author}{\bibfnamefont{F.}~\bibnamefont{{Schmidt-Kaler et~al}}},
  \bibinfo{journal}{J. Phys. B} \textbf{\bibinfo{volume}{36}},
  \bibinfo{pages}{623} (\bibinfo{year}{2003}).

\end{thebibliography}


\end{document}